\documentclass[11pt]{article}{}
\pdfoutput=1

\newcommand{\doctitle}{Inference of weak-form partial differential equations describing migration and proliferation mechanisms in wound healing experiments on cancer cells}
\newcommand{\docauthor}{P. Kinnunen, S. Srivastava, Z. Wang, ?? \& K. Garikipati}
\newcommand{\docaffil}{Applied Physics, Mathematics, Mechanical Engineering, and Michigan Institute for Computational Discovery \& Engineering\\ University of Michigan}

\newcommand{\docheader}{}
\newcommand{\docfooter}{University of Michigan}

\newcommand{\pdftitle}{Wound healing}
\usepackage{main}

\usepackage{xpatch}
\xapptocmd\appendices{%
  \crefalias{section}{appendix}%
}{}{\PatchFailed}
\crefname{section}{Sec.}{Secs.}
\Crefname{section}{Sec.}{Secs.}
\crefname{equation}{Eq.}{Eqs.}
\Crefname{equation}{Eq.}{Eqs.}
\creflabelformat{equation}{#2(#1)#3}
\crefname{figure}{Fig.}{Figs.}
\Crefname{figure}{Fig.}{Figs.}
\creflabelformat{figure}{#2#1#3}
\crefname{table}{Tab.}{Tabs.}
\Crefname{table}{Tab.}{Tabs.}
\creflabelformat{table}{#2#1#3}
\Crefname{appendix}{App.}{Apps.}
\crefname{appendix}{App.}{Apps.}
\Crefname{si}{Supp Inf.}{Supp Inf.}
\crefname{si}{Supp Inf.}{Supp Inf.}

\begin{document}

\pretitle{\begin{center}\vskip -80pt}%
\title{\Large\doctitle}
\posttitle{\end{center}}
\preauthor{\begin{center} \vskip -0pt}
\author[0,1]{Patrick C. Kinnunen}
\author[0,2,7,8]{Siddhartha Srivastava}
\author[2,7]{Zhenlin Wang}
\author[3]{Kenneth K.Y. Ho}
\author[3]{Brock A. Humphries}
\author[3]{Siyi Chen}
\author[1,4]{Jennifer J. Linderman}
\author[3,4,5]{Gary D. Luker}
\author[3,5]{Kathryn E. Luker}
\author[2,6,7,8]{Krishna Garikipati\thanks{Corresponding author at: Department of Aerospace \& Mechanical Engineering, University of Southern California, United States. \emph{E-mail address}: krishna.garikipati@usc.edu (K. Garikipati).}} 
\affil[1]{Department of Chemical Engineering, University of Michigan, United States}
\affil[2]{Department of Mechanical Engineering, University of Michigan, United States}
\affil[3]{Department of Radiology, University of Michigan, United States}
\affil[4]{Department of Biomedical Engineering, University of Michigan, United States}
\affil[5]{Biointerfaces Institute, University of Michigan, United States}
\affil[6]{Department of Mathematics, University of Michigan, United States}
\affil[7]{Michigan Institute for Computational Discovery \& Engineering, University of Michigan, United States}
\affil[8]{Department of Aerospace \& Mechanical Engineering, University of Southern California, United States}
\affil[0]{Equal contribution}
\postauthor{\end{center} \vskip -20pt}
\predate{\begin{center} \vskip -0pt}
\date{} 
\postdate{\end{center} \vskip -20pt}%
\maketitle
\begin{abstract}
\pagestyle{abstract}
Targeting signaling pathways that drive cancer cell migration or proliferation is a common therapeutic approach. A popular experimental technique, the scratch assay, measures the migration and proliferation--driven cell closure of a defect in a confluent cell monolayer. These assays do not measure dynamic effects. To improve analysis of scratch assays, we combine high-throughput scratch assays,  video microscopy, and system identification to infer partial differential equation (PDE) models of cell migration and proliferation. We capture the evolution of cell density fields over time using live cell microscopy and automated image processing. We employ weak form-based system identification techniques for cell density dynamics modeled with first-order kinetics of advection-diffusion-reaction systems. We present a comparison of our methods to results obtained using traditional inference approaches on previously analyzed 1-dimensional scratch assay data. We demonstrate the application of this pipeline on high throughput 2-dimensional scratch assays and find that low levels of trametinib inhibit wound closure primarily by decreasing random cell migration by approximately $20\%$. Our integrated experimental and computational pipeline can be adapted for  quantitatively inferring the effect of biological perturbations on cell migration and proliferation in various cell lines. 
\end{abstract}

\newpage
\section{Introduction}
Cell migration is a complex, multiscale phenomenon that integrates many different inputs and cell behaviors, including directed motion and random motion, that are differentially regulated by signaling kinases, cell density, and other factors\cite{mayor_front_2016, sengupta_principles_2021}. Cell migration helps maintain or form tissues or monolayers both in vivo and in vitro\cite{hiratsuka2015intercellular, scarpa2016collective}; cell division also contributes to the development or maintenance of these multicellular structures. Further, cell migration is associated with cancer metastasis, and thus is commonly studied in the context of oncogenic mutations or cancer treatments. 
Inferring relationships between perturbations and migratory outputs is challenging. Drugs could affect one or more different drivers of cell migration, potentially in different ways, with difficult-to-predict overall effects on cell migration. Cell migration is commonly measured using scratch assays, where a monolayer of cells is scratched to physically remove cells in a localized region\cite{liang_vitro_2007}. Subsequently, migration into the newly emptied space is measured over time. These measurements are commonplace in biological research, and have, for instance, been used to identify novel inhibitors of cell migration or genes responsible for regulating migration\cite{simpson2008identification, yarrow2005screening}. However, typically the complexity of cell migration is reduced to a single number, which represents the amount of space filled in a given time or the distance traveled by the leading edge of cells. This reduction to a single number obscures the behavioral changes that modify migration and could, for instance, fail to identify differences between drugs that inhibit migration to the same degree, but through the modulation of either directed migration or cell proliferation. Thus, the utility of scratch assays could be improved by identifying relationships between perturbations and specific biological effects.

Computational and mathematical modeling of cell migration has been used to extract granular, quantitative information from scratch assays. First-principles modeling has established a variety of partial differential equation (PDE) models that accurately represent the behavior of scratch assays with a high density of cells and under a variety of conditions\cite{jin_reproducibility_2016, maini_traveling_2004}. These models commonly include reaction terms, which represent cell proliferation or death, and diffusion terms, which represent random cell motion, with various functional forms. For example, some models use a reaction equation corresponding to cells growing with a constant proliferation rate, affected by a maximum carrying capacity. Other terms, including advection (directed cell migration), could also be included in cases such as chemotaxis or migration to regions of lower cell density where cells have a directional movement stimulus. By estimating parameters for these models, past work has helped identify quantitative effects of biological perturbations in scratch assays\cite{jin_reproducibility_2016, lagergren_biologically-informed_2020, gnerucci2020scratch}. However, models need to be flexible enough to account for behaviors observed in novel experimental conditions. 

Recent work has also established the use of physics-informed neural networks to model cell migration in scratch assays\cite{lagergren_biologically-informed_2020}. In this approach, a neural network is trained to predict the progression of cell concentrations over time. During training, the neural network optimizes a loss function which penalizes both inaccurate predictions and deviations from a pre-defined reaction-diffusion model. In this way, the neural network learns density-dependent proliferation and diffusion terms which best fit the data. However, in such approaches, neural networks can learn  relationships to make predictions without any guarantee that the growth or diffusivities learned are physically realizable. 

Another method for learning governing PDEs from experimental data and known physical constraints is Variational System Identification. It extends the popular SINDy approach \cite{KutzPNAS2015,KutzIEEE2016,KutzSCIADV2017,KutzChaos2018,KutzSIAM2019} to inferring PDEs in weak form, thereby offering advantages in the presence of noise and high derivatives \cite{wang_variational_2019, wang_system_2021, wang2021variational}. Variational System Identification enables modelers to identify a library of physically meaningful operators (e.g, differential terms such as the gradient, Laplacian and biharmonic operators, as well as algebraic ones such as polynomials, trigonometric functions and exponential functions) which make up the PDE. Variational System Identification then identifies parsimonious models incorporating a subset of operators by iteratively dropping terms from the library. Variational System Identification and SINDy respectively use the weak and strong forms of differential equations in regression-based approaches, and therefore do not require repeated forward evaluations of the model, unlike traditional PDE-constrained inverse modeling. This substantially reduces the computational cost. Variational System Identification has been applied to identify governing equations for the evolution of materials \cite{wang_variational_2019,wang2020perspective,wang2021variational}, the spatiotemporal spread of COVID-19\cite{wang2020system,wang_system_2021}, as well as constitutive models of soft materials \cite{wang2021inference,nikolov2022ogden}. After identifying a sparse set of operators, the coefficients of these operators in the inferred PDE can be fine-tuned using adjoint-based gradient optimization and validated against additional experimental test data. Finally, the surviving operators and parameters can be compared among models inferred from datasets under different conditions in order to extract quantitative insights from data. 

We hypothesized that Variational System Identification could be applied as above to wound healing data, eventually providing quantitative comparisons between different conditions. To test this hypothesis, we first applied the approach to previously published wound healing data \cite{jin_reproducibility_2016}. With it, we were able to rapidly identify an accurate model describing the evolution of cell concentration over time for a variety of initial seeding densities with accuracy that matched or improved upon results in the literature \cite{jin_reproducibility_2016}. Reaction and diffusion terms were identified underlying the observed behavior and consistent with previous results. We next applied Variational System Identification to our own data. We performed scratch assays on MDA-MB-231 breast cancer cells under a variety of conditions and used automated image processing to extract cell concentration fields over time. We found that Variational System Identification could successfully identify mechanisms of advection (directed motion), diffusion (random motion) and proliferation of cells in 2-D scratch assays by inferring the governing PDEs. Extending the approach's applications we quantitatively inferred the effect of trametinib, a MAPK pathway inhibitor, on cell diffusion and proliferation. We present Variational System Identification as a useful modeling approach to rapidly identify plausible models for cell population dynamics and quantify model effects under different experimental conditions. 

\section{Methods}
\subsection{Cell Culture}
We cultured MDA-MB-231 breast cancer cells in Dulbecco’s Modified Eagle Medium (DMEM) with $10\%$ fetal bovine serum (FBS). We passaged cells at a 1 to 10 ratio when they were approximately $90\%$ confluent. For imaging experiments, we cultured cells in imaging media, consisting of fluorobrite phenol red-free medium supplemented variable FBS (depending on desired experimental conditions), 1X penicillin/streptomycin, 1X GlutaMAX, and 1X sodium pyruvate. Sodium pyruvate was added as an antioxidant to reduce imaging-induced stress. 

\subsection{Stable Cell Line Generation}

We engineered the MDA-MB-231 cells used in this work to  express a stable histone-2B (H2B) nuclear marker (mCherry), as described previously\cite{spinosa2021pre, spinosa2019short}.

\subsection{Scratch Assay and live cell microscopy}
For scratch assay experiments, we seeded 50,000 MDA-MB-231 pHAEP cells in 1 ml imaging media in a 24-well glass-bottom imaging plate. We grew cells to full confluency (approximately 36 hours) before starting the scratch assay. For the scratch assay, we manually scratched each cell monolayer using a p200 pipette tip. Immediately after scratching, we washed the cells with 1 ml warm phosphate buffered saline (PBS) and then added 1 ml of warm imaging media containing experimental treatments.  

We imaged cells using an EVOS M7000 fluorescent microscope with on-stage incubator. After scratching each well, we placed the well plate into the prepared EVOS incubator. The incubator was maintained at 37°C, $5\%$ CO$_2$, and $>80\%$ humidity. For time-lapse imaging, we captured  fields centered on the scratch near the center of each well. We captured fluorescence from mCherry to position individual cells over time. We acquired images every 20 minutes over 24-48 hours in all wells in the well plate. We imaged one region of the scratch per well. 

\subsection{Automated image processing}\label{sec:auromated_image_processing}
We processed fluorescent images of the MDA-MB-231 pHAEP cells as described previously\cite{kinnunen2022computational}. Image processing was performed using MATLAB. Briefly, we first thresholded the nuclear images using an adaptive thresholding method. After identifying nuclear pixels, we extracted the centroid pixel of each distinct nuclear object. In contrast to previous work, we did not track each cell between frames. After automated image processing, we extracted cell density fields $C(\boldsymbol{x}, t)$ (number$/\mu\mathrm{m}^2$) from each well in the experiment. (Here, $\boldsymbol{x} = (x_1,x_2) \in \mathbb{R}^2$ is the two-dimensional position vector of a point.) To do so, we applied spatial binning to the cell positions. Bin sizes between 50$\mu\mathrm{m}$ and 100$\mu\mathrm{m}$ were used. We smoothed $C$ in  space and time. 
We applied a moving average filter with a window of 150 $\mu\mathrm{m}$ to spatial data and a window of 3 hour and 40 minutes to temporal data. 
Thus smoothing the data alleviates numerical noise in the calculation of derivatives. 

\subsection{Wound healing quantification}

We quantified wound closure by identifying the distance between wound edges at the first, $d(t_\text{start})$ and last $d(t_\text{end})$ time instants in the experiment. By aligning the wound edges with the $x_2-$axis, we estimated the position of the front of each wound as a line $x_1 = d(t)$ at time $t$ and calculated the distance in the $x_1$ direction (perpendicular to the wound) between each wound front. Then, we calculated wound closure as:

\begin{align}
    \text{Wound closure} = \frac{d(t_{\text{start}}) - d(t_{\text{end}})}{d(t_{\text{start}})}    
\end{align}

\section{Continuum-scale data-driven modeling for cell migration}

The density of cells is defined as a spatiotemporal field, $C(\bx,t)$ with $(\bx, t) \in \Omega \times [0,T]$ where $\Omega \subset \mathbb{R}^2$ and $[0,T]$ are the spatial domain and time period of interest. The evolution of this field is described using the following PDE, which is of first-order in time: 
\begin{equation}
    \frac{\partial C}{\partial t} = \mathscr{L}(C ; \boldsymbol{\theta})
\end{equation}

\noindent where, $\mathscr{L}$ is a differential operator parameterized by $\boldsymbol{\theta}$. Given $\boldsymbol{\theta}$ the PDE also can be stated in terms of the residual:
\begin{align}
    \mathscr{R}(C;\boldsymbol{\theta})&= \frac{\partial C}{\partial t} - \mathscr{L}(C ; \boldsymbol{\theta}).
    \label{eq:residual}
\end{align}
For some $\widetilde{C}$ that is not the solution of the PDE, $\mathscr{R}(\widetilde{C};\boldsymbol{\theta})$ is the pointwise ``error'' in satisfaction of the PDE.  To obtain the solution of the forward PDE problem means finding $C$ such that $\mathscr{R}(C;\boldsymbol{\theta}) = 0$, given $\boldsymbol{\theta}$. System identification, on the other hand, is an inverse problem in which, given data $C^\mathrm{d}(\boldsymbol{x}_i,t_j)$ at a finite number of sampling points $\boldsymbol{x}_i$ and times $t_j$ from measurements of a field representing $C$, we seek the optimal parameters $\theta^\ast$ satisfying 
\begin{align}
    \boldsymbol{\theta}^* = \argmin_{\boldsymbol{\theta} \in \boldsymbol{\Theta}} \left\vert\!\left\vert\!\left\vert \overline{\mathscr{R}}(\boldsymbol{\theta}| C^\mathrm{d}) \right\vert\!\right\vert\!\right\vert,\quad\text{where}\quad \overline{\mathscr{R}}(\boldsymbol{\theta}| C^d) = \mathscr{R}(C^\mathrm{d};\boldsymbol{\theta})
\end{align}
where the set of admissible parameters is $\boldsymbol{\Theta}$ and $\vert\!\vert\!\vert\bullet\vert\!\vert\!\vert$ is a suitable norm. The above statement of the system identification problem also holds for nonlinear PDEs. This approach to system identification reduces to a regression problem that is fairly inexpensive to solve computationally. It allows us to start with a large  library of candidate terms that could comprise $\mathscr{L}$ and subsequently eliminate the insignificant ones using principled approaches including regularization and stepwise regression.
Here, we have intentionally omitted  details such as the choice of norm $\vert\!\vert\!\vert\bullet\vert\!\vert\!\vert$, and functional spaces for $C^\text{d}$, in favor of presenting the basic idea of system identification. Finite element methods arrive at  the residual formulated in terms of the weak form of the PDE, instead of the strong form as in (\ref{eq:residual}). We adopt the weak form for PDE inference, and therefore refer to it as Variational System Identification \cite{wang_variational_2019, wang_system_2021, wang2021variational}. This technique is described next.

\subsection{Variational formulation of the advection diffusion reaction problem}

The advection-diffusion-reaction equation represents transport of cell density, and is encoded in the operator, $\mathscr{L}$. Mechanistically, the advection represents the directed motion of the cells, diffusion represents their random motion and reaction models cell proliferation/death. The advection-diffusion-reaction equation is written as follows: 
\begin{align}\label{eq:adr_strong}
    \frac{\partial C}{\partial t} =\nabla \cdot \left(D\nabla C\right)-\nabla \cdot(C v_{f}\bv_\text{unit})+r
\end{align}
where $D$ is the diffusivity, $v_f$ is the advective speed, $\bv_{unit}$ is a unit vector perpendicular to the wound, and $r$ is the cell reaction rate. The statement of the Initial and Boundary Value Problem (IBVP) includes the initial density $C(\boldsymbol{x},0)$, Dirichlet and Neumann boundary conditions on $\partial\Omega_\text{D}$ and $\partial\Omega_\text{N}$, respectively, where $\overline{\partial\Omega_\text{D}\cup\partial\Omega_\text{N}} = \overline{\partial\Omega}$ and $\partial\Omega_\text{D}\cap\partial\Omega_\text{N} = \emptyset$:
\begin{align}
        C(\bx,0) &= C_0(\bx), \quad \forall \quad \bx\in \Omega \nonumber\\
    C(\bx,t) &= \overline{C}(\bx, t), \quad \forall \quad \bx\in \partial\Omega_D, t\in [0,T]\nonumber\\
    \left( D \nabla C - C v_f \bv_\text{unit} \right) \cdot \bn &= q(\bx, t) , \quad \forall \quad \bx\in \partial\Omega_N, t\in [0,T]
\end{align}

\noindent The Dirichlet boundary condition specifies the cell density, and the Neumann boundary condition specifies the cell flux on the respective boundaries.

There exist models that have considered density-dependent effects on migration, for instance representing cells interacting with each other, and on cell proliferation, for instance, modeling its saturation at high cell density \cite{jin_reproducibility_2016, vittadello2019mathematical}. We incorporate such mechanisms in our model by considering parameters that are functions of cell density, $D(C)$, $v_f(C)$, and $r(C)$.  The weak form of \cref{eq:adr_strong} is:
\begin{align}\label{eq:adr_weak}
    \int_\Omega w\frac{\partial C}{\partial t} \mathrm{d}V
    &=\int_\Omega  \left( - D \nabla w \cdot \nabla C +  C v_{f}\nabla w\cdot\bv_\text{unit} + w r \right) \mathrm{d}V + \int_{\partial \Omega_\text{N}} wq   \mathrm{d}s
\end{align}
 for all $w$ in a suitable functional space, which we detail below. Here, $\bn$ is the unit outward normal vector on the boundary $\partial \Omega$. We seek the solution $C(\bx, t)$ in the Sobolov space $ H^1(\Omega)$, which consists of functions that are square-integrable and have square-integrable first derivatives.  The field $w$ is a weighting function and belongs to a subspace of $H^1(\Omega)$ defined as $\{w\in H^1(\Omega)\ | w=0 \text{ on } \partial\Omega_D \}$. 

\subsection{The finite element form}

The weak form of the advection-diffusion-reaction equation, given in \cref{eq:adr_weak}, can be discretized using the finite element method (FEM). The domain $\Omega$ is partitioned into $n_\text{el}$ elements, $\Omega = \cup_{e=1}^{n_\text{el}} \Omega_e$. The unknown field $C(\bx,t)$ is replaced by a finite-dimensional approximation $C^h(\bx,t)$, defined over each element using a linear combination of basis functions:

\begin{align}\label{eq:fe_approximation}
    C^h(\bx,t) = \sum_{i=1}^{n_\text{basis}} d_i(t) N_i(\bx)
\end{align}
where $d_i(t) = C^\text{d}(\boldsymbol{x}_i,t)$ are the time-dependent coefficients at finite element nodes $\boldsymbol{x}_i$, and the spatial dependence of $C^h(\boldsymbol{x},t)$ is represented by the finite element basis functions $N_i(\bx)$ (also known as shape functions). The span of these basis functions defines a finite dimensional $H^1$ space: $V^h = \text{span}(\{N_i\}_{i=1}^{N_\text{{basis}}}) \subset H^1({\Omega})$. There are many ways to define these basis functions but for our study we will choose piecewise linear basis functions for one- and two-dimensional problems, the latter using triangular elements \cite{hughes2003finite}.
Also within the Galerkin approach, we consider finite dimensional weighting functions, $w^h\in \{V^h(\Omega) | w^{h} = 0 \text{ on } \partial \Omega_D\}$ 
which we write as $w^h(\boldsymbol{x}) = \sum_{j=1}^{n_\text{wt}}b_jN_j$ where $N_j$ are the $n_\text{wt} \le n_\text{basis}$ basis functions of $V^h$ that vanish on $\partial\Omega_D$. Substituting these finite-dimensional approximations $C^h$ and $w^h$ for $C$ and $w$, respectively, in (\ref{eq:adr_weak}) and invoking its validity for all $w^h \in V^h$ we obtain the following residual vector: 
\begin{align}\label{eq:adr_residue}
    \bR(C^h;\boldsymbol{\theta})
    =&\int_\Omega \left( \frac{\partial C^h}{\partial t} \bN d\Omega
    + D \nabla \bN \cdot \nabla C^h 
    -  C^h v_{f}\bv_\text{unit}\cdot \nabla \bN 
     - r \bN \right) d\Omega 
    \nonumber\\
    &- \int_{\partial \Omega_T} \bN\left( D  \nabla C^h - C^h v_{f}\bv_\text{unit}\right).\bn   ds
\end{align}
where $\bN$ is the vector of finite element basis functions, and the second integral on the right imposes the Neumann boundary condition. The residual vector $\boldsymbol{R}(C^h;\boldsymbol{\theta})$ is a finite-dimensional version of $\mathscr{R}(C;\boldsymbol{\theta})$. 




\subsection{Inference of the advection diffusion reaction system}

We infer the advection-diffusion-reaction system by posing it as a standard optimization problem using the residual derived in the previous section and the cell density field from wound healing experiments (see \cref{sec:auromated_image_processing}). The extracted density field is written by interpolation over finite element nodal values as in \cref{eq:fe_approximation}. We impose Dirichlet boundary conditions on all edges,. Therefore, the Neumann boundary integral of \cref{eq:adr_residue} does not feature in the following discussion.

We consider the following ansatz for the $D$, $v_f$ and $r$ parameters:
 \begin{subequations}
     \begin{align}
	D&= \theta_0\cdot 1 + \theta_1 C + \theta_2 C^2\\
	v_f&= \theta_3\cdot 1 + \theta_4 C + \theta_5 C^2\\
	r&= \theta_6 C + \theta_7 C^2
 \end{align}
  \label{eq:ansatz}
 \end{subequations}

\noindent where $C$ is replaced by $C^h$ from \cref{eq:fe_approximation} during inference. The constant reaction term was omitted, since it represents cell proliferation ($r>0$) or death ($r<0$) in a region with zero cell density, both of which are unphysical. We can now rewrite the residual equations equation as a matrix-vector problem with the following form, where $\asemtext$ represents the finite element assembly operator. 

\begin{align}
    \bR(C^h; \boldsymbol{\theta} ) &= \by - \boldsymbol{\Xi}\cdot \boldsymbol{\theta}\\
    \by &= \asem\limits_e \left[-\int_{\Omega_e}\frac{\partial C^h}{\partial t}\bN d\Omega\right] \\
    \boldsymbol{\Xi} &\equiv \left[\boldsymbol{\Xi}_{0}, \cdots ,\boldsymbol{\Xi}_{7}\right].\
\end{align}
Here, $\boldsymbol{N} = \{N_k\}$ represents the vector of basis functions for $k = 1,\dots n_\text{np}$ where $n_\text{np}$ is the number of nodes in the problem, and $\boldsymbol{\Xi}$ is a matrix where the columns $\boldsymbol{\Xi}_j$ represent the residual vector terms corresponding to diffusive, advective and reaction operators in weak form from \cref{eq:adr_residue}. The rows correspond to the components of the basis function vector $\boldsymbol{N}$ concatenated over timesteps. We have,
\begin{align}
    \left[\boldsymbol{\Xi}_{0}\; \boldsymbol{\Xi}_{1}\; \boldsymbol{\Xi}_{2}\right] &= \asem\limits_e \left[\int_{\Omega_e}\left[1\; C^h\; C^{h^2} \right] \nabla \bN\cdot \nabla C^h \mathrm{d}\Omega \right]\\
    \left[\boldsymbol{\Xi}_{3}\; \boldsymbol{\Xi}_{4}\; \boldsymbol{\Xi}_{5} \right] &= \asem\limits_e \left[-\int_{\Omega_e}\left[1\; C^h\; C^{h^2} \right] C^hv_f\bv_{\text{unit}} \cdot \nabla\bN  \mathrm{d}\Omega \right]\\
    \left[\boldsymbol{\Xi}_{6}\; \boldsymbol{\Xi}_{7}\right] &= \asem\limits_e \left[\int_{\Omega_e}\left[ C^h\; C^{h^2} \right]\bN \mathrm{d}\Omega\right]    
\end{align}
We finally state the inference problem for the optimal parameters as the following minimization problem with a quadratic cost defined as the Euclidean norm $\vert\boldsymbol{R}\vert$:
\begin{align}\label{eq:vsi_regression}
    \boldsymbol{\theta}^* = \argmin_\theta \vert\bR(C^h; \boldsymbol{\theta})\vert^2 
\end{align}	

In the Variational System Identification framework, the maximum order of derivative evaluated on the data field is first order (\cref{eq:vsi_regression}), despite originating from a model based on a second-order differential equation. This results from employing the weak form of the PDE in our inference process. In the weak form, the derivatives that would traditionally act on the trial solution are instead transferred to the weighting functions. Evaluating derivatives on the data field can potentially amplify noise; however, limiting the derivative order to the first rather than the second, as is typical in strong form-based approaches, helps mitigate this issue to a significant extent. This choice enhances the robustness of the inference against noise in the data field.


In the interest of model parsimony, we seek to estimate the most significant terms given   ansatz \cref{eq:ansatz} for the parameters using stepwise regression \cite{wang_variational_2019, wang_system_2021, wang2021variational}. Since an increase in parsimony comes at the cost of an attendant growth in the loss between model iterates, we adopt a search across surviving terms and  select for elimination the candidate that, when excluded from the basis, leads to the minimal growth in the loss of the reduced optimization problem from \cref{eq:vsi_regression}. This results in a parameter vector $\overline{\boldsymbol{\theta}}$ that is sparse in the sense that most of its components $\overline{\theta}_m = 0$ for $m \in \{0,\dots,7 \}$.


    \subsection{PDE Constrained Optimization for model refinement} 

Following the inference of a parsimonious model that fits the data, PDE-constrained optimization with adjoint-based gradient computation is applied to improve the fit. The optimization problem statement is:

\begin{align}\label{eq:adjoint_loss}
\text{Find } \boldsymbol{\theta}^{\ast\ast} = \argmin_{\boldsymbol{\overline{\theta}}} \ell(\boldsymbol{\overline{\theta}};C^{\prime^h}), \;\text{where } \ell(\boldsymbol{\overline{\theta}}) &= ||C^{\prime^h}(\boldsymbol{x},t;\boldsymbol{\overline{\theta}}) - C^h(\boldsymbol{x},t)||_{L_2 (\Omega \times [0, T])}^{2}\nonumber\\
\text{such that } \boldsymbol{R}(C^{\prime^h};\boldsymbol{\overline{\theta}}) &= \boldsymbol{0} \nonumber\\
\text{for } C^{\prime^h}(\boldsymbol{x},t) &= \sum_{i=1}^{n_\text{basis}} d_i^\prime(t) N_i(\bx)\quad\text{defined over each element}  
\end{align}

\noindent Here, $C^{\prime^h}$ is a finite element interpolant field that is different from the data-derived field $C^h$ introduced in equation \cref{eq:fe_approximation}. Thus,  the optimization problem in (\ref{eq:adjoint_loss}) involves solution of  the forward model, which can be computationally expensive and require exploring regions of parameter space that are numerically unstable, making it unsuitable for large models. However, if  a parsimonious model has been identified, then the approach in (\ref{eq:adjoint_loss}) can be applied to refine the PDE parameters starting from their values $\overline{\boldsymbol{\theta}}$ inferred by Variational System Identification as initial iterates. We solve the forward model in the second line of (\ref{eq:adjoint_loss})  using experimentally observed initial conditions, generating a predicted cell density field, which enables computation of an associated loss in the first line of (\ref{eq:adjoint_loss}). However that minimization problem requires computing a variational derivative through the chain rule which also involves evaluating the derivative of $C^{\prime^h}$ with respect to $\overline{\boldsymbol{\theta}}$. We complete this step via adjoint-based gradient optimization with the BFGS solver. PDE-constrained optimization with adjoint-based gradient computation does not change the selected operators but generates a new set of parameter values,   which  fit the data better than the those obtained by Variational System Identification with stepwise regression.

\subsection{Computational framework for inference}
The numerical examples presented in this work have been posed and solved in two dimensions by the finite element method programmed using the  \texttt{FEniCS} platform\cite{dolfin, logg2012automated}. The 1D forward solutions presented in this work were carried out on a uniform mesh with 42 piecewise linear elements. The 2D solutions were obtained on a rectangular domain using a structured mesh with 6478 linear triangular elements. The nonlinear optimization for minimizing \cref{eq:vsi_regression} was carried out using the \textit{Newton solver} in the \texttt{FEniCS} package. The PDE constrained optimization \cref{eq:adjoint_loss} was implemented in the \texttt{dolfin-adjoint} package\cite{dolfin-adjoint, dolfin-adjoint-optimization}.

\section{Results}

\subsection{Variational system identification of 1-D wound healing dynamics from data}
We first tested Variational System Identification  on a previously reported scratch assay dataset. The dataset, collected by Jin et al.,\cite{jin_reproducibility_2016} consists of scratch assays performed on PC-3 prostate cancer cells with varying initial confluence prior to the scratch. Different confluencies were achieved by varying the initial  seeding over 10,000 to 20,000 cells in steps of 2000. The high cell densities that  were achieved in this experiment justified the consideration of continuum advection-diffusion-reaction PDEs to model these data. For each of the resulting six datasets, cell density  was measured every 12hr for 48hr and averaged across three replicates. Next, the density was averaged along the ($x_2$) direction parallel to the scratch, yielding a one-dimensional concentration field (along $x_1$) for each initial density. 

    \begin{figure}
    \centering
    \includegraphics[width = 0.99\linewidth]{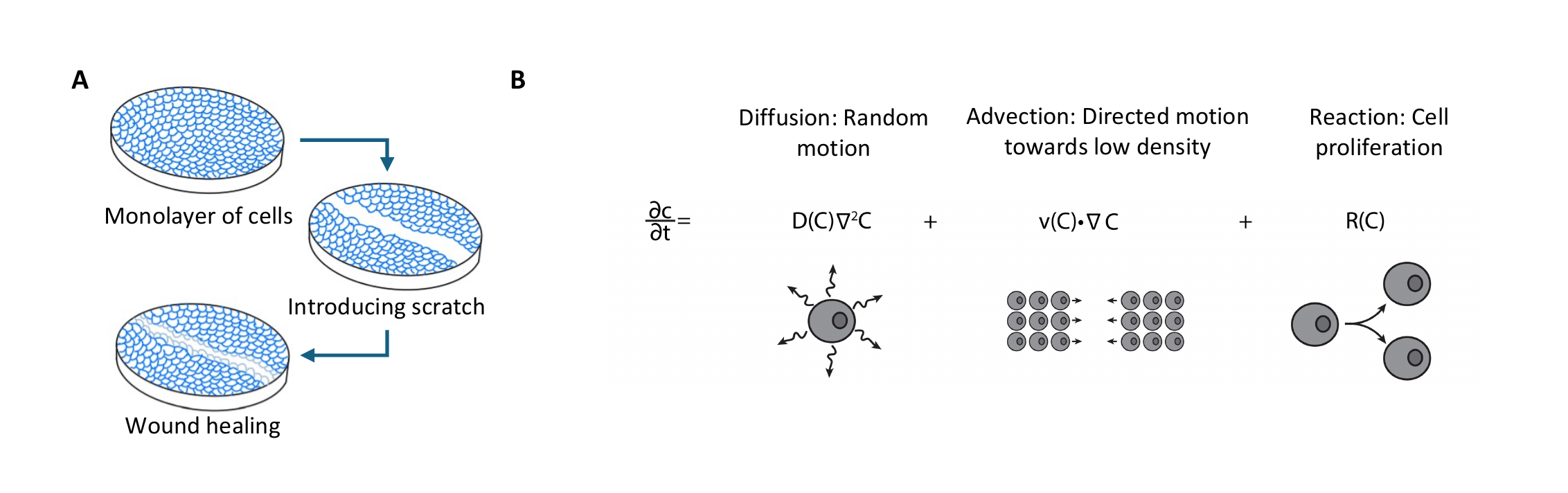}
    \caption{A) Illustration of wound healing/scratch assay. B) Diffusion, advection, and reaction mechanisms for cell random motion, directed motion and proliferation,r respectively. }
    \label{fig:benchmark_1d}
\end{figure}

\begin{figure}
    \centering
    \includegraphics[width = 0.85\linewidth]{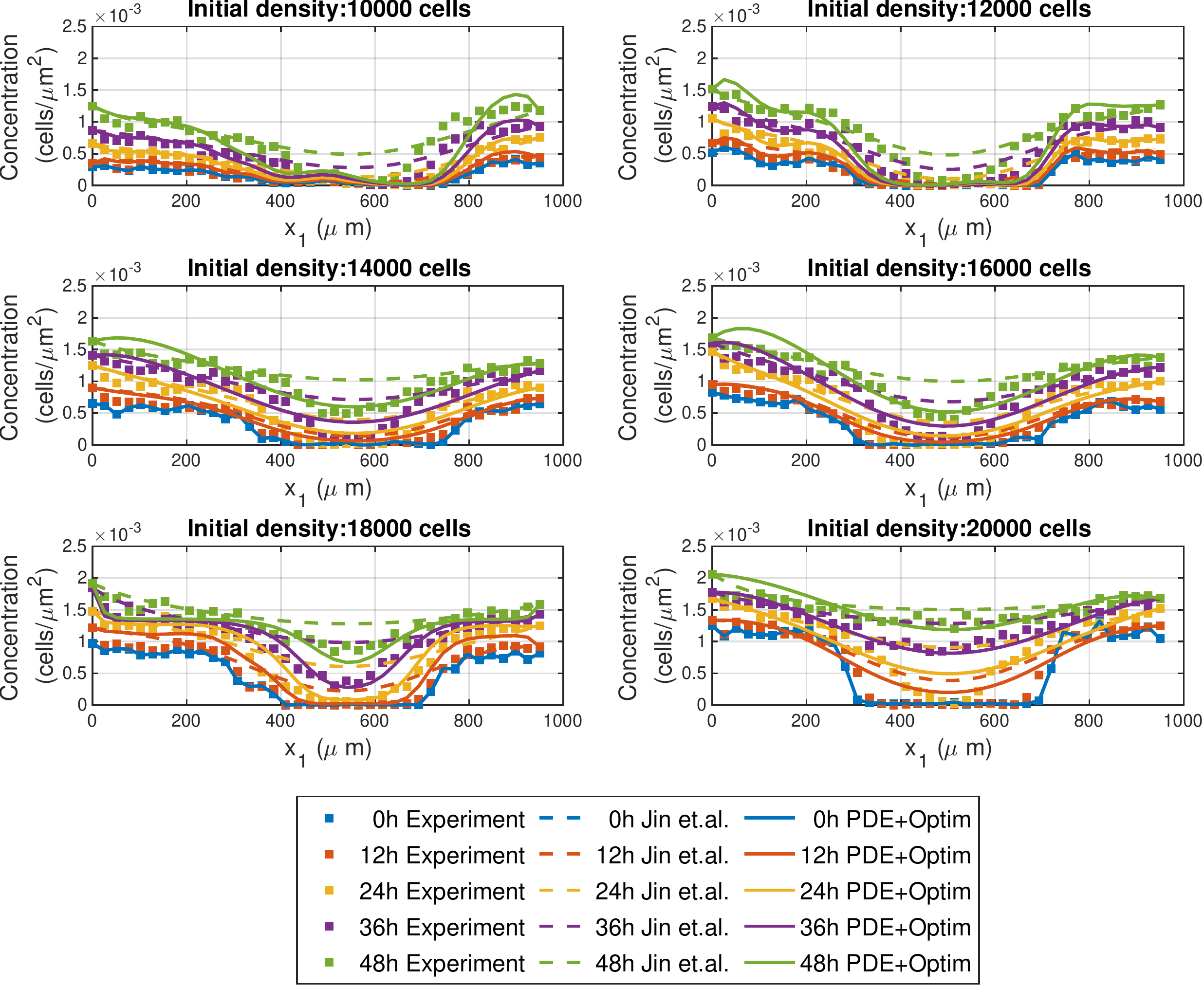}
    \caption{Experimental wound healing data (symbols), the model inferred by Jin et al.\cite{jin_reproducibility_2016} (dashed curves), and the model inferred by Variational System Identification and refined by PDE-constrained optimization (solid curves).}
    \label{fig:benchmark_results}
\end{figure}

The 1D cell density profiles over time are shown in \cref{fig:benchmark_results}. The scratch, which is approximately $450 \mu\mathrm{m}$ long, is clear in the initial density profiles. For all initial seeding with 14000 and more cells, the center of the scratch (around $x_1 = 500 \mu\mathrm{m}$) is occupied by cells after 36 hours. Furthermore, there is an increase in cell density at the scratch edge (near $x = 0 \mu\mathrm{m} \text{ or } 950 \mu\mathrm{m}$) over time for all initial seeding densities, indicating that the cells continue growing during the experiment.

As a proof of concept, we applied Variational System Identification to each of the six scratch assay datasets corresponding to different initial cell densities. From the data for $C^\mathrm{d}(x_1,t)$ we generated diffusivity, advective velocity, and reaction rate terms that are polynomial functions of concentration, $C^\mathrm{d}$ up to second order (quadratic \cref{eq:ansatz}).

\begin{figure}
    \centering
    \includegraphics[width = 0.7\linewidth]{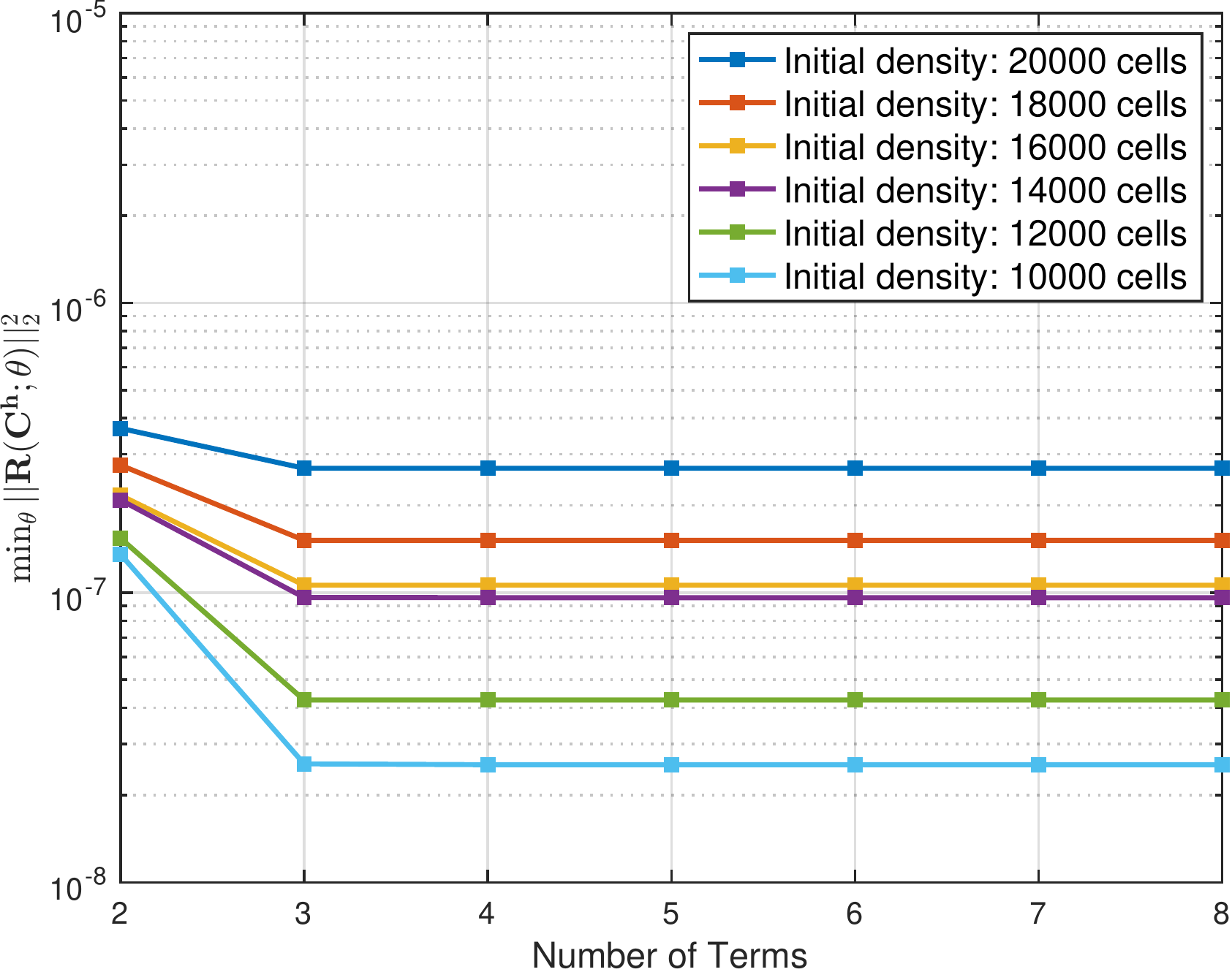}
    \caption{Variational System Identification loss calculated as the mean squared error between the calculated time derivatives and predicted derivatives from each stage of stepwise regression during Variational System Identification.}
    \label{fig:benchmark_vsi}
\end{figure}

\begin{figure}
    \centering
    \includegraphics[width = 0.7\linewidth]{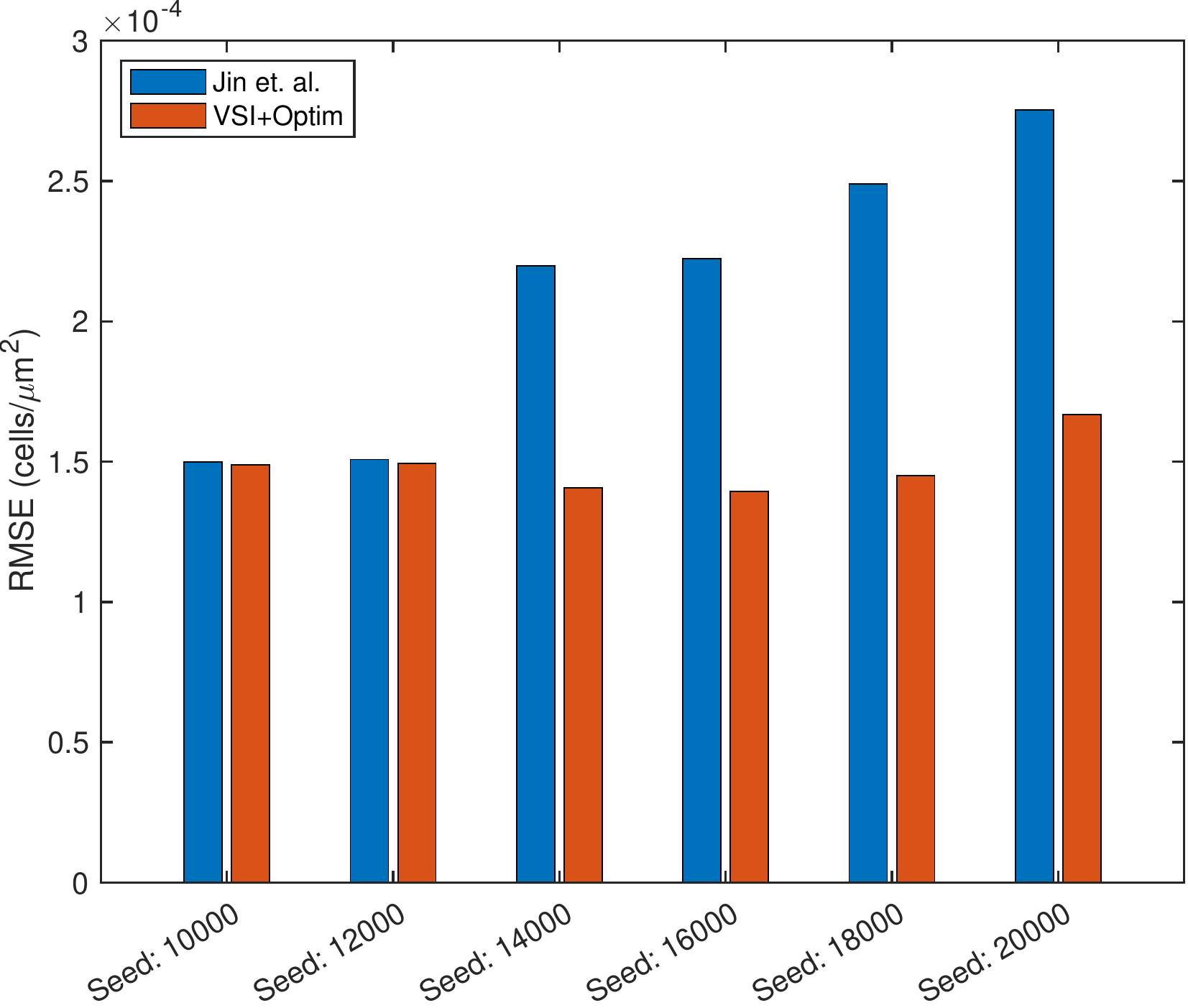}
    \caption{Root mean squared error (RMSE) calculated between experimental data and reaction-diffusion models from Jin et. al (blue), and a model obtained by PDE-constrained optimization following Variational System Identification  (red) for each case with different initial number of seeded cells.}
    \label{fig:benchmark_rmse}
\end{figure}

We next employed Variational System Identification to find parsimonious models. \cref{fig:benchmark_vsi} shows that for all cases labelled by the initial number of cells seeded, the loss (mean squared error between data and model predicted time derivatives) does not decrease for models having more than 3 terms. In every case, the three-term model excluded advection. We next performed PDE-constrained parameter optimization of the 3-term model for each of the six cases with different initial number of cells seeded, with the additional physical constraint of positive diffusivities. All the resulting optimized models have a constant diffusivity term, and either a  linear or quadratic dependence on concentration. All the reaction terms have linear concentration dependence, except  the four-term model for 18,000 cells (see \cref{tab:benchmark_results}). {\color{black} For the case of 18,000 cells, we observed minimal improvement in the  cost function \cref{eq:adjoint_loss} from the PDE-constrained optimization. The model, both before and after refinement, is presented in \cref{tab:benchmark_results} under the label 18,000 (3 terms). Furthermore, we noticed that in the 3-term model, the partial derivatives of the cost function with respect to the diffusivity coefficient remained low, indicating that the available data was not sufficiently sensitive to the diffusive mechanism of random cell motion. To address this issue, we expanded the model to include four terms—incorporating both linear and quadratic reaction terms. This expansion led to a significant reduction in the cost function during the PDE-constrained optimization. The refined model, presented in \cref{tab:benchmark_results} under the label 18,000 (4 terms), demonstrates a strong agreement between the model solution and the experimental data.} We used these refined models to run a forward simulation from the initial conditions of each dataset seeded with the different numbers of initial cells seeded. The model predictions and experimental observations appear in \cref{fig:benchmark_results}. We also present the forward simulation of the reaction-diffusion model using the parameters provided by Jin et.al.\cite{jin_reproducibility_2016}. We draw the reader's attention to the qualitative match between the forward solutions and the dynamics observed in the dataset at early times and the quantitative match  at later times. We found that for four of the six initial seeding densities, our models have root mean squared errors (RMSEs) that are significantly lower than the RMSE reported by Jin et al. and for two of the six cases they are similar (\cref{fig:benchmark_rmse}). 
These results suggest that our Variational System Identification and PDE-constrained parameter optimization modeling approach can be used to infer models for cell migration dynamics that are competitive with previous methods such as those by Jin et al, which fit a model based on prior knowledge about the system. {\color{black} In this work, we considered a polynomial cell concentration dependence in the diffusivity for modeling nonlinear diffusion, but at higher cell densities,  more rigorous models, such as Maxwell-Stefan diffusion would be more appropriate \cite{srivastava2024pattern}.}

\begin{table}[]
\begin{tabular}{|c|c|cl|c|}
\hline
Initial density        & Method       & \multicolumn{2}{c|}{Diffusivity ($\mu m^2/hr$)}           & Reaction Term ($1/hr$) \\ \hline
\multirow{2}{*}{10000} & VSI inferred & \multicolumn{2}{c|}{$7.9 + 9.0\times10^3 C$}             & $3.6\times10^{-2} C$      \\ \cline{2-5} 
                       & VSI+Optim    & \multicolumn{2}{c|}{$7.9 + 8.9\times10^3 C$}             & $2.7\times10^{-2} C$      \\ \hline    
\multirow{2}{*}{12000} & VSI inferred & \multicolumn{2}{c|}{$15. + 2.8\times10^3 C$}             & $2.7\times10^{-2} C$      \\ \cline{2-5} 
                       & VSI+Optim    & \multicolumn{2}{c|}{$15. + 2.9\times10^3 C$}             & $2.3\times10^{-2} C$      \\ \hline 
\multirow{2}{*}{14000} & VSI inferred & \multicolumn{2}{c|}{$17. + 7.1\times10^3 C$}             & $2.1\times10^{-2} C$      \\ \cline{2-5} 
                       & VSI+Optim    & \multicolumn{2}{c|}{$4.6\times10^2 + 0.25 C$}            & $2.3\times10^{-2} C$      \\ \hline 
\multirow{2}{*}{16000} & VSI inferred & \multicolumn{2}{c|}{$13. + 1.3\times10^2 C^2$}           & $1.9\times10^{-2} C$      \\ \cline{2-5} 
                       & VSI+Optim    & \multicolumn{2}{c|}{$3.6\times10^2 + 1.3\times10^2 C^2$}      & $2.3\times10^{-2} C$      \\ \hline     
\multirow{2}{*}{18000 (3 Term)} & VSI inferred & \multicolumn{2}{c|}{$17. + 1.3\times10^{-3} C$}          & $1.7\times10^{-2} C$      \\ \cline{2-5} 
                       & VSI+Optim    & \multicolumn{2}{c|}{$17. + 1.3\times10^{-3} C$}          & $1.5\times10^{-2} C$      \\ \hline    
\multirow{2}{*}{18000 (4 Term)} & VSI inferred & \multicolumn{2}{c|}{$17. + 1.3\times10^{-3} C$}          & $1.7\times10^{-2} C + 1.0\times10^{-8} C$      \\ \cline{2-5} 
                       & VSI+Optim    & \multicolumn{2}{c|}{$36.4 + 1.2\times10^{-2} C$}          & $9.5\times10^{-2}C  -69.7 C^2$      \\ \hline    
\multirow{2}{*}{20000} & VSI inferred & \multicolumn{2}{c|}{$22. + 3.1\times10^{-2} C^2$}        & $1.2\times10^{-2} C$      \\ \cline{2-5} 
                       & VSI+Optim    & \multicolumn{2}{c|}{$8.1\times10^2 + 3.1\times10^{-2} C^2$}   & $1.7\times10^{-2} C$      \\ \hline
\end{tabular}
\caption{Model terms learnt by Variational System Identification and PDE-constrained parameter optimization for each experimental condition presented in Jin et.al.. $C$ is the cell number density measured in the units of cells/$\mu m^2$. }
\label{tab:benchmark_results}
\end{table}

\subsection{Quantification of the effect of different levels of an inhibitor on migration dynamics in high-throughput, 2-dimensional scratch assays}

\begin{figure}
    \centering
    \includegraphics[width = 0.45\linewidth]{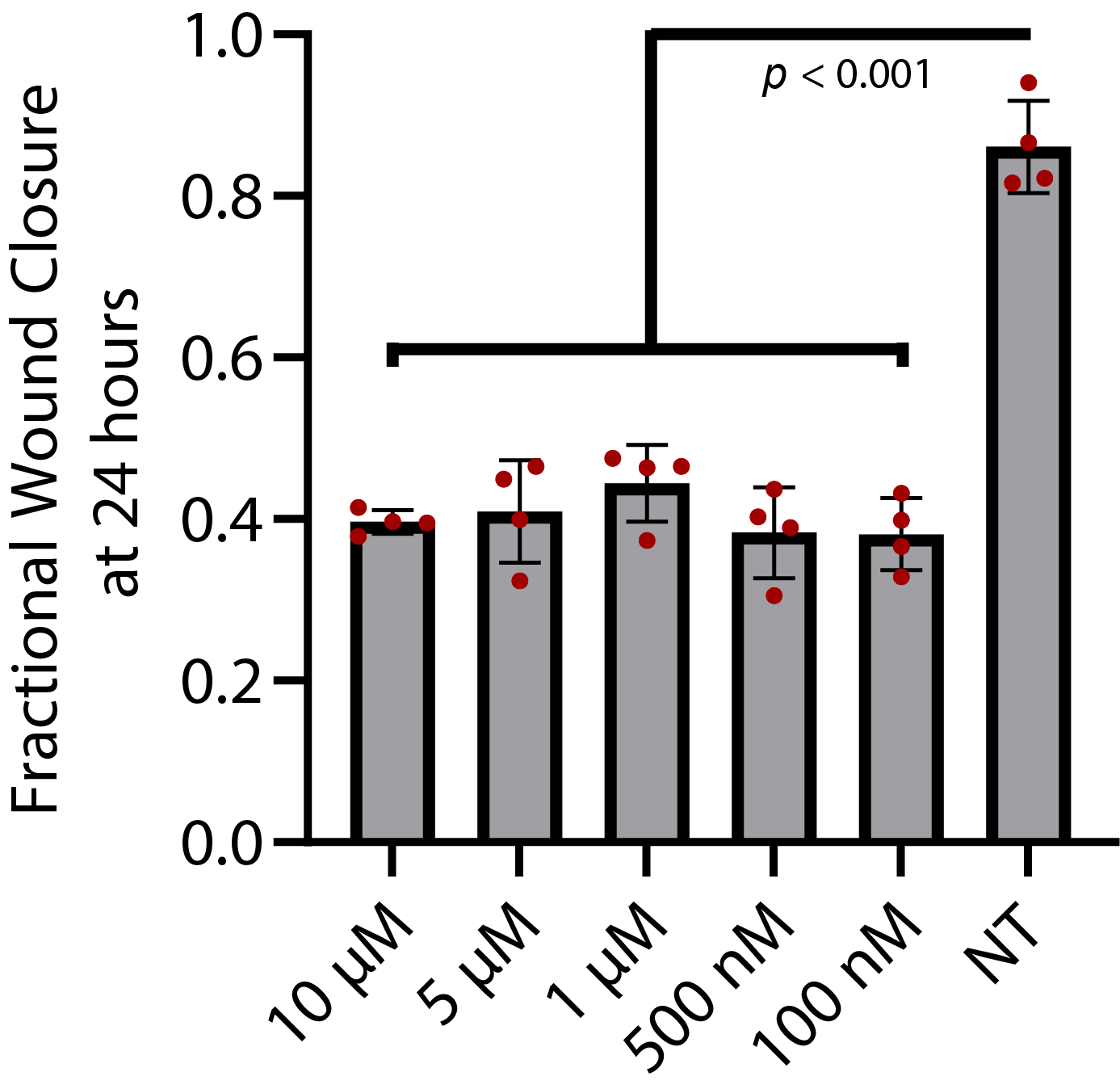}
    \caption{Fractional wound closure (Distance at 24 hr./initial distance) calculated for a range of trametinib concentrations and an untreated control, with $N = 4$ wells for all conditions.}
    \label{fig:fractional_wound_closure}
\end{figure}

\begin{figure}[H]
    \centering
    \includegraphics[width = 0.9\linewidth]{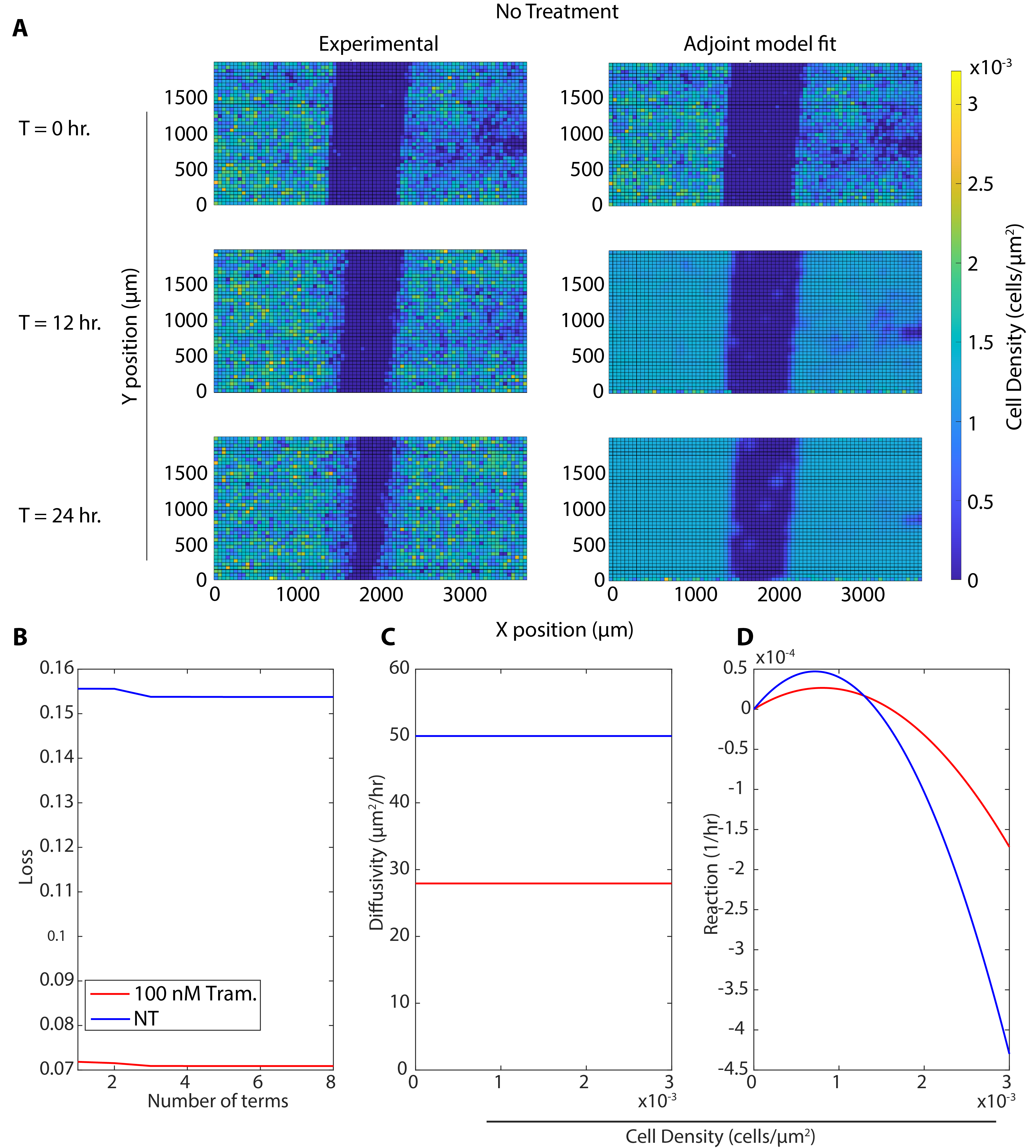}
    \caption{A) Comparison of experimental data and  predictions of a model learned by Variational System Identification followed by PDE-constrained parameter optimization  under the same initial conditions. The experimental condition is with no treatment. B) Variational System Identification loss as a function of the number of  terms in the model inferred for the untreated and 100 nM trametinib conditions. (C-D)  PDE-constrained and optimized diffusivity (C) and reaction  (cell proliferation) (D) terms as functions of concentration. Diffusivity is constant, while reaction/proliferation is a combination of linear and quadratic terms. The horizontal axis ranges were chosen to represent the concentrations present in the experimental data.}
    \label{fig:wound_healing_2d}
\end{figure}

After demonstrating that Variational System Identification and PDE-constrained optimization could be used to learn parsimonious advection-diffusion-reaction models for cell migration data in the literature, we used the approach  on our own cell density data gathered from high-throughput fluorescence microscopy experiments. We performed scratch assays using MDA-MB-231 breast cancer cells and tracked scratch closure over time using live-cell fluorescence microscopy \cref{fig:fractional_wound_closure}. As seen in \cref{fig:wound_healing_2d}A, the cell densities achieved are comparable to those observed by Jin et al.,\cite{jin_reproducibility_2016} and admit treatment of the problem using continuum advection-diffusion-reaction PDEs. 

We used this assay to compare how trametinib, a MEK kinase inhibitor, affects cell migration. 
MEK kinase is known to regulate cell migration and proliferation \cite{akinleye2013mek}. Our past study showed that trametinib reduces both random and directed motion in chemotaxis \cite{ho2023oscillatory}. We captured wound-closure dynamics for cells exposed to 5 concentrations of trametinib from 10 $\mu$ M to 100 nM, and for an untreated control. We observed that untreated cells were able to almost close the wound after 24 hours, while 100 nM trametinib was enough to significantly inhibit cell migration. Higher concentrations of trametinib did not further inhibit cell migration (\cref{fig:fractional_wound_closure}).

\begin{figure}[H]
        \centering    \includegraphics[width=0.75\linewidth, trim={0 7cm 0 6cm}]{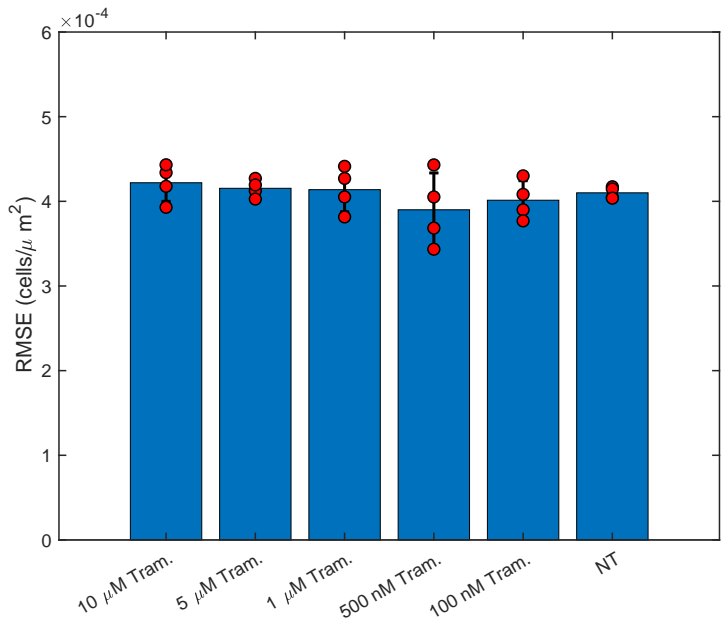}
    \caption{RMSE evaluated between the forward prediction of VSI+Optim models (Variational System Identification and PDE-constrained parameter optimization) presented in \cref{tab:adj_model_compare}. The red dots represent the RMSE for each replicate and the bar reperesents the mean value.}
    \label{fig:rmse_tram}
\end{figure}

We next applied our PDE inference pipeline to learn models for the observed wound closure dynamics. We used the same library of candidate operators, including constant, linear, and quadratic terms for diffusion and advection, and linear and quadratic terms for reaction. We fit a model to the replicate conditions simultaneously, generating a single model for each condition. Finally, we ran the inferred model for each condition starting from the experimentally measured initial cell density (\cref{fig:wound_healing_2d}A). We observed that the Variational System Identification loss function is essentially constant for two-term models including diffusion and reaction terms, and only starts to increase for models that lack reaction terms (\cref{fig:wound_healing_2d}B). Based on the these results, we adopted a three-term model. For both the untreated condition and 100 nM trametinib, these terms corresponded to constant (concentration-independent) diffusivity, and reaction terms that are linear and quadratic in cell density. 
\cref{tab:adj_model_compare} shows the coefficients inferred initially by Variational System Identification in the process of delineating mechanisms, and their subsequent refinement by PDE-constrained optimization. 
This combination of techniques reveals a non-linear functional dependence of cell proliferation (reaction in the PDE) rate on cell density: positive at low cell densities and  becoming negative above a critical density; i.e., cell death sets in(\cref{fig:wound_healing_2d}D).  This inferred mechanism could be interpreted as a crowding-induced induced death in the cell population. The models inferred with and without trametinib suggest that even 100 nM trametinib causes a decrease in random cell migration and cell proliferation. Notably, the effect of trametinib treatment quickly saturates, as is reflected in the coefficients for constant diffusivity and linear/quadratic reaction terms in  \cref{tab:adj_model_compare}. The RMSEs between the forward predictions of the model and the data are presented in \cref{fig:rmse_tram}, and are essentially comparable across trametinib concentrations and for the untreated case. The mean value across conditions is $\sim 13\%$ of the maximum cell concentrations in the experiments shown in \cref{fig:wound_healing_2d}A. 

\begin{table}[]
\begin{tabular}{|c|c|c|c|c|}
\hline
\multicolumn{1}{|c|}{Condition} & \multicolumn{1}{c|}{Method} & \multicolumn{1}{c|}{Diffusivity ($\mu m^2/hr$)} & \multicolumn{1}{c|}{Reaction Term ($1/hr$)}  \\ \hline
\multirow{2}{*}{10 $ \mu $M}    & VSI inferred  & $6.8$   & $5.5\times10^{-2}C  -3.4\times10^1C^2$ \\ \cline{2-4} 
                                & VSI+Optim     & $2.7\times10^1$ & $8.2\times10^{-2}C  -5.2\times10^1C^2$ \\ \hline
\multirow{2}{*}{5 $ \mu $M}     & VSI inferred  & $7.2$   & $5.0\times10^{-2}C  -3.0\times10^1C^2$ \\ \cline{2-4} 
                                & VSI+Optim     & $2.8\times10^1$   & $7.0\times10^{-2}C  -4.4\times10^1C^2$ \\ \hline
\multirow{2}{*}{1 $ \mu $M}     & VSI inferred  & $6.2$   & $5.2\times10^{-2}C  -3.2\times10^1C^2$ \\ \cline{2-4} 
                                & VSI+Optim     & $2.3\times10^1$   & $7.1\times10^{-2}C  -4.5\times10^1C^2$ \\ \hline
\multirow{2}{*}{500nM}          & VSI inferred  & $7.8$   & $4.8\times10^{-2}C  -2.9\times10^1C^2$ \\ \cline{2-4} 
                                & VSI+Optim     & $3.0\times10^1$   & $6.3\times10^{-2}C  -4.0\times10^1C^2$ \\ \hline
\multirow{2}{*}{100 nM}         & VSI inferred  & $7.3$   & $5.1\times10^{-2}C  -3.0\times10^1C^2$ \\ \cline{2-4} 
                                & VSI+Optim     & $2.8\times10^1$   & $6.6\times10^{-2}C  -4.1\times10^1C^2$ \\ \hline
\multirow{2}{*}{NT}             & VSI inferred  & $9.1$   & $9.1\times10^{-2}C  -6.0\times10^1C^2$ \\ \cline{2-4} 
                                & VSI+Optim     & $5.0\times10^1$   & $1.3\times10^{-1}C  -9.2\times10^1C^2$ \\ \hline
\end{tabular}
\caption{Model terms learnt by Variational System Identification and PDE-constrained parameter optimization for each experimental condition shown in Figure 3. $C$ is the cell number density measured in the units of cells/$\mu m^2$}
\label{tab:adj_model_compare}
\end{table}

\section{Discussion}

The treatment of cell migration via continuum PDEs is well established in biophysics \cite{narayanan2010silico,garikipati2017perspectives, jennifer_book}, as is the role of cell arrangement in determining aspects of the progression of cancers \cite{mills2014elastic}. Here, we demonstrate that our PDE inference approach of Variational System Identification followed by PDE-constrained parameter optimization can infer parsimonious, quantitative and physics-based models for collective cell migration. In a 1D setting, we infer models with  accuracies that are competitive with other recent approaches. 
We also have applied our approach to our own 2D wound healing experiments. Relative to the 1D data, our 2D wound healing experiments were sampled more frequently (every 20 minutes compared to every 12 hours) with similar spatial resolution. In this setting, we estimate diffusivity values for breast cancer cells around 50 $\mu m^2/hr$ in untreated conditions, with decreasing diffusivity in response to trametinib. Previous computational analyses of cell migration have identified a wide range of diffusivity values, from approximately 50 to 3000 $\mu m^2/hr$\cite{maini_traveling_2004, sengers2007experimental, cai2007multi}. Thus, our estimates of diffusivity in the 1D and 2D settings are consistent with the prior literature. 

There are some advantages to  inference  by Variational System Identification followed by PDE-constrained parameter optimization compared to other approaches such as traditional model inference or Physics Informed Neural Networks (PINNs). Traditionally, building models of cell migration has utilized iterative cycles of model conceptualization, based on known cell biology, that include increasingly complex models for cell division or diffusion. These models are informative and have made substantial progress towards a general theory of collective cell migration. However, model development can require a substantial amount of expertise or trial and error. Furthermore, this approach does not scale well if new sources of data are acquired, which could 
increase the number of potential model terms based on the interaction of a new data source with all other types of data previously considered in the model. For example, data about cell state (e.g. signaling activity, cell-cycle status, metabolic activity) gathered from fluorescent reporters or cell morphology could require updating a model to consider diffusivity as a function of both cell state and local cell density\cite{hiratsuka2015intercellular, gordonov2016time, vittadello2020examining, aoki2017propagating}. Hence, we envision that Variational System Identification and PDE-constrained parameter optimization can improve traditional cell dynamics and signaling modeling in at least two ways. First, it could serve as a hypothesis-testing tool. When considering a library of mathematically expressed candidate mechanisms and datasets to test against, modelers could use the approach advanced here to develop models that explain the data within specified error bounds and with a desired parsimony. Second, the approach presented here could identify new experimental conditions to generate data that activate mechanisms not queried by existing datasets. Traditional modeling can be time consuming, and a more rapid, semi-automated model testing approach such as Variational System Identification combined with PDE-constrained parameter optimization could thus lead to more tightly coupled model-driven experimentation and data generation.

Other modeling approaches have used neural networks to perform data-driven inference on reaction and diffusion equations governing cell migration. In Variational System Identification, the equations learned are confined to the candidate library terms used, while neural networks can learn arbitrary relationships for equation parameters to best fit the data. Lagergren et al. demonstrated the power of this approach on the Jin et al. dataset\cite{jin_reproducibility_2016} (also analyzed by Jin et al.), when they inferred density-dependent reaction-diffusion equations for cell migration\cite{lagergren_biologically-informed_2020}. They identified non-linear functions of concentration for diffusivity and reaction terms, and found that these functions vary with initial seeding concentration, consistent with our findings. There is one noteworthy difference between neural network-based approaches and Variational System Identification. In the case of cell migration, while the neural networks can learn arbitrary relationships between cell density and diffusivity or reaction, there is no guarantee that  physically realizable diffusivity or reaction relationships are learnt, unless a number of physics-based constraints are built in. Variational System Identification, on the other hand, enables the modeler to restrict the candidate mechanisms and their mathematical forms to those that rigorously encode physical mechanisms. PINNs can discover  relationships that maximize the fit to data, while Variational System Identification sacrifices fitting to data because it is constrained to selected terms which the modeler can ensure are physically meaningful. A possible criticism of the approach is that inference proceeds by model selection from a library of candidates, and not by a method of \emph{de novo} discovery.
 
We envision two key applications of Variational System Identification followed by PDE-constrained parameter optimization furthering our understanding of biology. First, it can be used to quantify the effects of drugs in high-throughput screens. Scratch assays have been miniaturized and mechanized, making them compatible with high-throughput screening\cite{poon2017device, yue2010simplified}. Thus, the approach presented here with trametinib could be extended and combined with high-throughput drug screens so that the specific effects of drugs on cell migration and division could be determined. Second, our approach could be used to rapidly infer models based on new streams of data gathered from scratch assays. Cell morphology\cite{gordonov2016time} and fluorescent reporters\cite{aoki2017propagating} have been used to measure or infer cell states in migrating cells. Thus, Variational System Identification followed by PDE-constrained parameter optimization could be used to resolve diffusivity or reaction terms as functions of not only cell density but also local measures of cell state. 

 \section{Conclusion}
 
In summary, we demonstrate that Variational System Identification followed by PDE-constrained parameter optimization can be used to infer  models of collective cell migration and proliferation/death in a wound healing assay. We benchmark this method by comparing inferred models with previous reaction-diffusion models applied to 1-D wound healing data, finding that our models are comparable or slightly superior in accuracy to the previously published, traditionally derived models. We next capture cell migration in 2-D wound healing assays using video microscopy, measuring the effect of serum on cell migration as a test case. We further find that the approach advanced here can be applied to 2-D wound healing to quantify effects of a targeted inhibitor on cell migration and density-dependent effects on proliferation. Our work demonstrates that this  pipeline can be used to rapidly identify parsimonious models for cell migration and proliferation  from scratch assays. In the future, we aim to address limitations in experiments that may not resolve length scales with sufficiently high cell densities for continuum-level predictions. In such cases, we plan to leverage the relationship between transport PDEs and Brownian motion-based stochastic differential equations to further inform the inference procedure when working with sparse data.

\section*{Acknowledgements}

The authors acknowledge funding from United States National Institutes of Health Grants R01CA238042,  R50CA221807 (K.E.L), R01CA238023, R33CA225549, R37CA222563, and R21AI173559. We also acknowledge support from the W.M.Keck Foundation.

\pagestyle{bibliography}
\bibliography{main}

\end{document}